\documentclass[10pt,twocolumn,letterpaper]{article}

\usepackage{cvpr}              

\definecolor{cvprblue}{rgb}{0.21,0.49,0.74}
\usepackage[pagebackref,breaklinks,colorlinks,allcolors=cvprblue]{hyperref}

\newcommand{\ours}[0]{\text{WISE}\xspace}

\usepackage{soul}
\usepackage{textcomp}
\usepackage{stfloats}
\usepackage{xcolor}
\usepackage{hyperref}
\usepackage{amsmath}
\usepackage{algorithm}
\usepackage{algorithmic}

\usepackage{amsthm,amsfonts,caption,subcaption,graphicx,stmaryrd,booktabs,arydshln,amssymb,multirow,bbold,enumerate,textcomp,float}
\usepackage{hhline,wrapfig}

\def\paperID{18374} 
\def\confName{CVPR}
\def\confYear{2025}

\title{WISE: A Framework for Gigapixel Whole-Slide-Image Lossless Compression}

\author{
Yu Mao\textsuperscript{1}, 
Jun Wang\textsuperscript{2}, Nan Guan\textsuperscript{2},  Chun Jason Xue\textsuperscript{1} \\
\textsuperscript{1}MBZUAI, \textsuperscript{2}City University of Hong Kong \\
{\tt\small \{yu.mao, jason.xue\}@mbzuai.ac.ae} 
{\tt\small \{jwang699-c, nanguan\}@cityu.edu.hk}
}

\begin{document}
\maketitle
\begin{abstract}
Whole-Slide Images (WSIs) have revolutionized medical analysis by presenting high-resolution images of the whole tissue slide. 
Despite avoiding the physical storage of the slides, WSIs require considerable data volume, which makes the storage and maintenance of WSI records costly and unsustainable. 
%
To this end, this work presents the first investigation of lossless compression of WSI images.
Interestingly, we find that most existing compression methods fail to compress the WSI images effectively. 
Furthermore, our analysis reveals that the failure of existing compressors is mainly due to information irregularity in WSI images.
To resolve this issue, we develop a simple yet effective lossless compressor called \ours, specifically designed for WSI images.
\ours employs a hierarchical encoding strategy to extract effective bits, reducing the entropy of the image and then adopting a dictionary-based method to handle the irregular frequency patterns.
%
%
Through extensive experiments, we show that \ours can effectively compress the gigapixel WSI images to $36$ times on average and up to $136$ times. 
\end{abstract}    
\section{Introduction}
\label{sec:intro}

\begin{figure}[t]
  \centering
\includegraphics[width=0.44\textwidth]{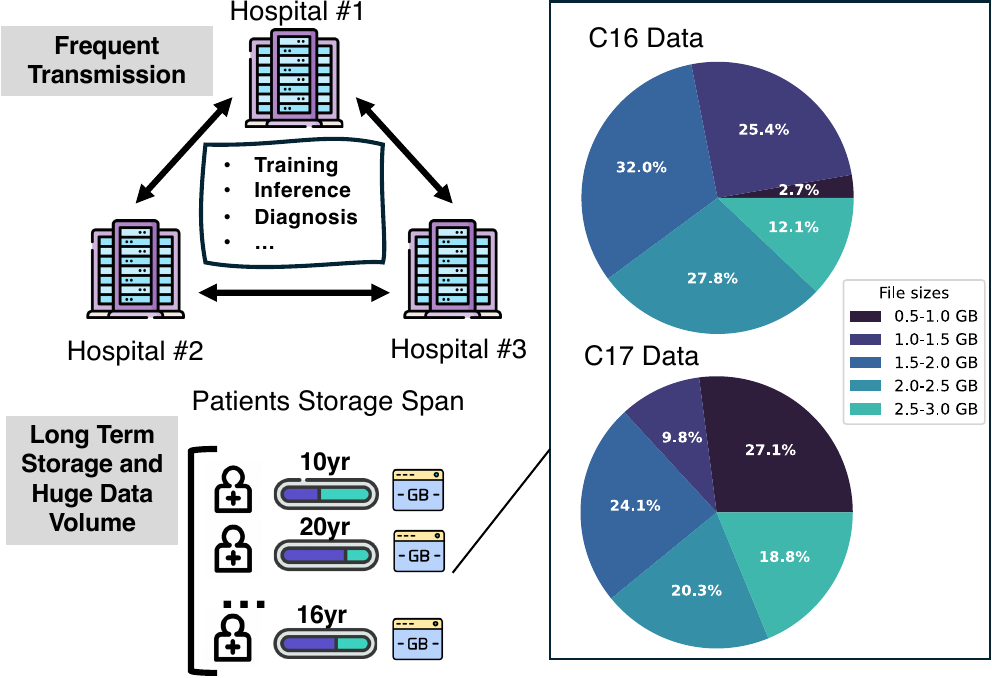}
  \caption{WSIs are extremely large images ($\sim$3 GB/slide uncompressed) of high resolution (0.25 microns per pixel), presenting a significant data storage challenge for hospitals wishing to adopt digital pathology. In practice, hospitals often ship physical drives via FedEx to transfer WSIs. Therefore, there is a desperate need for better compression of WSIs.}
  \label{fig:motivation}
\end{figure}

The ever-increasing utilization of advanced medical imaging techniques, such as Whole-Slide Images (WSI)~\citep{Kumar2020WholeSI, wang2024advances}, has led to a surge in the generation of gigapixel medical image data. 
These detailed volumetric images are indispensable for accurate diagnosis and effective treatment planning. However, their considerable size presents notable challenges for storage and transmission. 
%
For instance, a single digital pathology WSI scan can produce a four-dimensional (4D) image of a patient’s tissue sample comprising width, height, color channels, and multiple resolution levels, 
resulting in a total data volume reaching several gigabytes~\citep{keighley2023digital}.  
As shown in Fig.\ref{fig:motivation}, two well-known datasets, C16~\citep{DATASETc16} and C17~\cite{DATASETc17}, consist primarily of whole slide images (WSIs) ranging in size from 1 to 5GB. The large size of these images presents significant challenges for storage and transmission. Considering a person got a WSI image as a teenager, the WSI records need to be stored for decades to keep track. Those cold, huge data can cost a lot for storage. Also, as the cloud-based medical ecosystem expands, the frequency of WSI transfers between data centers is increasing due to the growing demands for training, inference, and diagnosis~\citep{liu2025diffkillr, DATASETc17, wang2024shap}.

Given the aforementioned challenges in using WSI images for digital pathology, there exists a clear need for efficient and lossless WSI compressors. 
%
However, most existing WSI compression approaches focus on lossy compression, including those classical methods that leverage lossy compressors such as JPEG-2000, as well as those modern compressors that utilize DNN to encode WSI patches into embeddings~\cite{keighley2023digital, helin2018optimized, faghani2023optimizing, barsi2024deep}.
While lossy compression techniques can achieve high compression ratios, they risk introducing distortions that could compromise the diagnostic integrity of the WSI images and potentially lead to medical issues. 
%
On the other hand, lossless compression techniques can avoid distorting the original WSI information. However, our preliminary experiments with existing lossless compression techniques show that even state-of-the-art lossless compressors can not yield the desired compression performance.
Traditional lossless image compressors like PNG, entropy encoders like Huffman, or dictionary compressors like Gzip, demonstrate limited effectiveness on WSI images. Even modern Neural Network (NN)-based compressors can achieve only limited compression, around 1.6 to 2 times for the non-empty area in WSI images, which is far from satisfactory given that a single WSI image contains tens of gigabytes.

Therefore, in this work, we fill this gap by developing an efficient lossless WSI compressor.
First, we observe that most WSI images consist of empty areas that can be efficiently encoded as simple coordinate tuples. 
While removing these empty regions can easily save some storage space, our findings indicate that the true challenge in losslessly compressing WSI images lies in handling their informative regions.
Following this observation, we conduct an in-depth analysis of the compression properties of WSI images, building on a preliminary evaluation of existing lossless compression methods.
%
Our analysis reveals that WSI images generically exhibit \textit{high information irregularity}, where the high-frequency signals are widely distributed across the WSI images, and demonstrate high volatility.
The irregular frequency patterns pose significant challenges to prevalent entropy-based or prediction-based compression methods.
%
%
Based on those findings, we propose a simple yet effective dictionary-based lossless compression method called \ours, specifically designed for WSI images.
%
\ours consists of three major steps: 
i) including a hierarchical projection coding step, 
ii) a bitmap coding step, and iii) a dictionary-based compressor.
Through i) and ii), \ours effectively reduces the entropy of the sequence and disentangles proper locality patterns for dictionary-based compression.
We conduct extensive experiments and demonstrate that by using our proposed \ours method, WSI images can be compressed to $36$ times on average and up to $136$ times smaller.
%
Our contributions can be summarized as follows:
\begin{itemize}[leftmargin=*]
    \item To the best of our knowledge, this paper is the first to conduct a comprehensive study on the application of various lossless methods in WSI;
    \item This paper reveals why lossless methods often fall short with WSI images and offers insights into enhancing the effectiveness of lossless compression for WSI applications;
    \item We propose a simple yet powerful lossless compression method called \ours based on a composite encoding scheme;
    \item We conduct extensive experiments that verify the effectiveness of \ours.
\end{itemize}


\section{Related Work}
%
%

\paragraph{Medical Image Compression} has attracted great interest in the community.
Several existing works have adopted \textit{lossy} compression techniques such as JPEG-2000 or VQVAE~\cite{keighley2023digital, helin2018optimized, faghani2023optimizing, barsi2024deep}, and compress WSI patches into compact embeddings~\citep{lazardself, tellez2020extending, chan2023histopathology}. Reconstruction from such embeddings has also gained popularity~\cite{zhang2021nonsmooth, zhuang2021geometrically}; however, these approaches inevitably suffer from information loss inherent in the compressed representations.
%

\paragraph{Patch-based Lossless Compression Approaches}
typically divide an image into multiple small patches, which are then compressed individually.
Typical patch-based compression methods include traditional image compression techniques, such as PNG~\cite{png} and TIFF~\cite{tiff}.
%
%
In recent years, there has been emerging interest in applying deep learning methods to image compression and it has intrigued multiple advanced compression technologies, such as
Bit-Swap~\cite{bitswap}, IDF~\cite{hoogeboom2019integer}, IDF++~\cite{idf++}, and ArIB-BPS~\cite{zhang2024learned}. 
However, deep learning methods are still based on entropy encoding and thus does not perform well on WSI slides.

\paragraph{Byte Stream-based Image Compression Approaches} treats the image as a byte stream input, onto which to perform the compression operations to reduce the storage requirements.
These methods can be further divided into entropy-based and dictionary-based methods.
%
%
Common entropy-based methods include Huffman coding~\cite{huffman1952method}, Arithmetic coding~\cite{arithmetic}, and PixelCNN~\cite{van2016pixelcnn,salimans2017pixelcnn++}. Huffman coding~\cite{huffman1952method} is a classical entropy-based coding method that assigns shorter codes to more frequently occurring symbols.
Arithmetic coding~\cite{arithmetic} is a more sophisticated method that assigns a unique floating-point range to the entire input sequence, often offering higher compression rates than Huffman coding. PixelCNN~\cite{van2016pixelcnn,salimans2017pixelcnn++} uses convolutional neural networks to model the conditional dependencies between pixels, enabling efficient compression by learning the image’s probability distribution. TRACE~\cite{mao2022trace} and PAC~\cite{mao2023faster} utilize lightweight transformers and MLP to model dependencies between adjacent pixels, respectively.
%
Common dictionary-based methods include Gzip~\cite{gzip}, LZMA~\cite{lz77}, and Zstandard~\cite{zstd}. 
Gzip~\cite{gzip} is widely used in file compression, which is built upon the LZ77~\cite{lzw} compression algorithm combined with Huffman coding for further optimization. 
LZMA~\cite{lz77} enhances efficiency by using bitfield-specific contexts to represent literals or phrases. 
Zstandard~\cite{zstd} combines a dictionary-matching stage (LZ77) with a large search window and a dual entropy-coding stage using Huffman coding for literals and Finite State Entropy (FSE) for sequences.
However, although these methods can achieve a certain amount of lossless compression ratio on WSI images, the compression ratio is not satisfied.
\begin{figure*}[t]
  \centering
\includegraphics[width=0.8\textwidth]{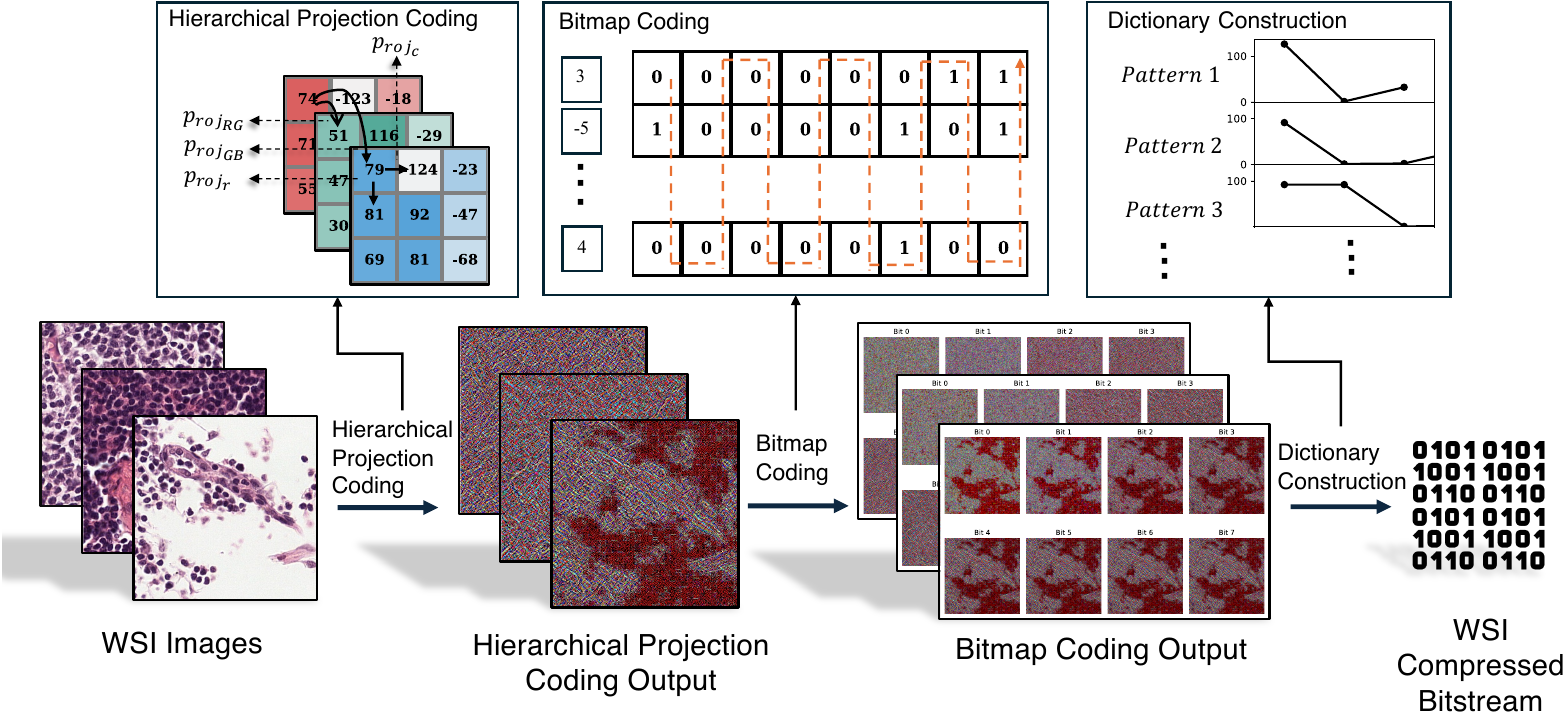}
  \caption{Illustration of the \ours pipeline. \ours compress the input WSI images through three key steps. As WSI images exhibit a high degree of information imbalance, the first step in \ours is effective information extraction, which performs image preprocessing, hierarchical projection encoding, and bitmap encoding, 
  to transform the input image into information-dense encodings. To be specific, effective bits are gathered together for better pattern extraction.
  Then, in the second step, \ours constructs a dictionary to capture the byte pattern in transformed image for further compression.}
  \label{fig:pipeline_overview}
\end{figure*}

\section{\ours Framework}
To overcome the information irregularity issue of the WSI image compression, inspired by the analysis in the previous section, we propose a dictionary-based compression method called \ours. In what follows, we first provide an overview of \ours, and then elaborate more on the details of each step in \ours. 

\subsection{Information Irregularity in WSI Images}
\label{sec:motivation}



Essentially, the success of many compression methods relies on identifying the locality properties within adjacent pixels or the pixel value distributions.
Nevertheless, WSI images may demonstrate a different locality pattern from normal images such that existing compression methods fail to achieve satisfactory performance.

\begin{figure}[t]
  \centering
\includegraphics[width=0.45\textwidth,angle=0]{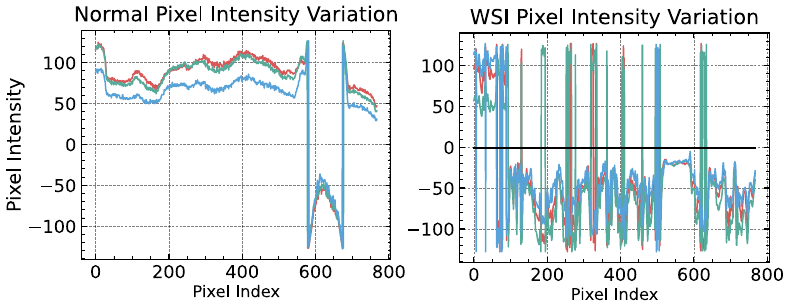}
  \caption{WSI image's pixel intensity variation is quite different from normal image. It has a large portion of high-frequency features thus affecting the performance of entropy-based compressors.}
\label{fig:freq_analysis}
\end{figure}

To begin with our analysis, we start by conducting an empirical study on the data extracted from a normal image from Kodak dataset~\cite{kodak} and a sample WSI image from TCGA~\cite{DATASETtcga}.
The results are given in Table~\ref{tab:empirical}. 
Specifically, we consider five prevalent lossless compression methods, covering all the mainstream compression approaches, 
including traditional compressors like Huffman~\citep{huffman1952method}, Pfor~\cite{zukowski2006superscalar}, Arithmetic Coding~\cite{arithmetic}, PNG~\citep{png}, and LZMA~\cite{lzw}. 
%
%
The compression ratio values reflect the effectiveness of each algorithm in reducing WSI image size, where higher values generically indicate better compression performance for the given data. 
%
Specifically, the best compressor for normal image data, PNG, can only achieve little-to-no compression ratio on WSI images. On the contrary, the dictionary-based method lzma yields a relatively high compression ratio on WSI images. 
The same phenomenon happens to the other compressors such as Huffman, Pfor, and PAC.
Nevertheless, the simple dictionary-based compression method LZMA can achieve relatively high compression ratios on both normal and WSI images. 

To better understand the failure reasons of the existing compression methods. We further analyze the characteristics of the WSI images. As the identification of locality properties has been shown to be quite sensitive to the frequency patterns in the input data~\citep{rhee2022lc}, we focus on analyzing the frequency characteristics of the WSI images.
%
As illustrated in Fig.~\ref{fig:freq_analysis}, compared with normal images, WSI images demonstrate unique properties in the frequency domain, which we term as \textit{information irregularity}.
Specifically, the frequency information in the WSI images consists of a large part of the high-frequency features and more local extremes. The \textit{high volatility} of the WSI data poses a significant challenge to capturing the locality for compression, especially for prediction-based compressors~\cite{faghani2023optimizing, rhee2022lc}.

Based on this analysis, we can better understand why dictionary-based methods, which typically perform poorly on multi-dimensional image data, outperform image-specific compressors like PNG on WSI images.
The reason is that constructing a dictionary instead of predicting the sequential patterns is more robust to the high volatility of the WSI data. 
Yet, a near 2x compression ratio remains insufficient for WSI data (often with sizes of several gigabytes). 


\begin{table}[t]
\centering
\caption{Empirical study for a normal image from Kodak datatet and a WSI Image from TCGA.}
\begin{tabular}{l|cccccc}
\hline
 & \textbf{Huffman}  & \textbf{Pfor} & \textbf{PAC} & \textbf{PNG} & \textbf{LZMA} \\
\hline
\textbf{Normal} & 1.19  & 1.42 & 1.98 & 2.06 & 1.43 \\
\textbf{WSI} & 1.23  & 0.99 & 0.87 & 1.01 & 1.93 \\
\hline
\end{tabular}
\label{tab:empirical}
\end{table}




\subsection{Pipeline Overview}
WSI images are typically stored in a multi-resolution pyramid structure. In this paper, we focus on compressing only the base level of the pyramid, as the other layers can be generated by downsampling this primary level. We process the WSI in a patch-based manner. 

The \ours framework consists of four steps, as shown in Fig.~\ref{fig:pipeline_overview}.
In the first step, we identify the non-zero regions in the WSI images, as it is common that a huge part of the WSI images contain meaningless information (i.e., pixels in zero value).
In the second step, we incorporate a hierarchical linear projection encoding algorithm to remove the unnecessary information between adjacent regions or pixels.
%
In the third step, we investigate the correlations at an even more fine-grained level to achieve a higher compression ratio. Specifically, we consider the \textit{correlations of the bits} from the same positions in each byte, and further compress the image reorganized via byte transformation.
Finally, with proper encodings obtained from the last two steps, we introduce a dictionary-based compression method that captures the local repetitive patterns. 


\subsection{Preprocessing}
%
A considerable part of sizable WSI images consists of regions outside the tissue area (so-called “white-space” without any clinically relevant information). In this study, we first developed an image-processing algorithm that can remove the unneeded background in a WSI. 
Specifically, we remove areas where the column sum or the row sum has the value of zero.
Another pre-process we apply is to remove the alpha channel. Different from normal image data, the original WSI images contain four channels, where there is additional alpha information. We directly ignore the additional alpha channel as it contains little to no information about the diagnosis.
%
We apply all the downstream operations to the remaining informative areas with three channels.



\begin{figure*}[t]
  \centering
\includegraphics[width=\textwidth,angle=0]{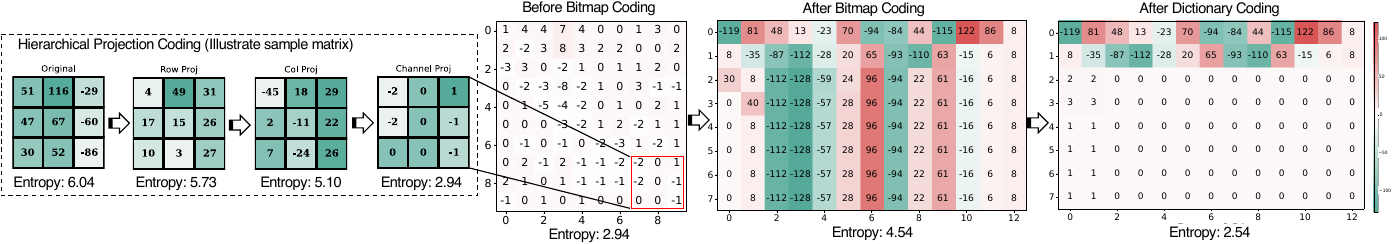}
  \caption{Detailed illustration of the Hierarchical Projection Coding, bitmap coding and dictionary coding on a sample matrix. Starting from the original information (left), Hierarchical Projection Coding performs row projection, column projection, and channel projection sequentially, which gradually reduces the entropy of the projected encodings. Then, the bitmap coding transposes the effective bits, coding the matrix with small values to a highly repetitive matrix, as shown in the bottom middle. This approach would increase the byte-based entropy since the repetitive byte is reduced. Nonetheless, after dictionary coding, the entropy would be further reduced. }
  \label{fig:hp_encoding}
\end{figure*}

\subsection{Hierarchical Projection Coding}
%

Since the WSI images demonstrate heterogeneous patterns in the frequency domains, we consider a hierarchical encoding structure of the whole image information to achieve maximal compression. 
Instead of directly storing the entire dataset, the method stores the first-order difference between consecutive data points across three dimensions: row, column, and channel. Each pixel has three reference data points, which are taken from the nearest points in its row, column, and channel, based on physical proximity to the target pixel. This approach is chosen because WSIs often exhibit significant variability, making it challenging to model over long distances. This variability is also the reason why prediction-based methods, as discussed earlier, tend to be ineffective.
The main purpose of this technique is to optimize storage space and reduce bandwidth usage by capturing and representing only the differences (or deltas) between sequential data points,
such that one could use a smaller regime of values to represent pixels.
The detailed algorithm is illustrated in Alg.~\ref{alg:delta}.

We provide a sample to illustrate the effectiveness of the proposed Hierarchical Projection Coding. Figure~\ref{fig:hp_encoding} presents a sample matrix extracted from an informative WSI area, showing its entropy after each step of the linear projection process. 
The entropy \( H \) of a matrix \( M \) can be calculated using Shannon's entropy formula~\cite{shannon1948mathematical}:

\[
H(M) = - \sum_{i=1}^{n} p(x_i) \log_2(p(x_i))
\]
where: \( x_i \) represents each unique value in the matrix, \( p(x_i) \) is the probability of \( x_i \) (i.e., the frequency of \( x_i \) divided by the total number of elements), \( n \) is the total number of unique values in the matrix.
As observed, with the application of each coding level, the entropy of the matrix progressively decreases. According to Shannon's information theory, the lower the entropy, the higher the compression ratio~\cite{shannon1948mathematical}. Evaluation of compression ratio on six WSI datasets will be provided in Sec.~\ref{sec:exp_comp_rate}.








\begin{algorithm}[htbp]
\caption{Hierarchical Projection Coding for WSI Image Patch}
\label{alg:delta}
\begin{algorithmic}[1]

\REQUIRE Input WSI image patch \( X \in \mathbb{R}^{M \times N \times C} \)
\ENSURE Processed image patch \( Y \in \mathbb{R}^{M \times N \times C} \)
\FOR{each channel \( c = 1 \dots C \)}
    \FOR{each pixel \( (m, n) \) for \( m = 2 \dots M \)}
        \STATE \( \Delta X_{m,n,c} \gets X_{m,n,c} - X_{m-1,n,c} \)
    \ENDFOR
    
    \FOR{each pixel \( (m, n) \) for \( n = 2 \dots N \)}
        \STATE \( \Delta^r X_{m,n,c} \gets \Delta X_{m,n,c} - \Delta X_{m,n-1,c} \)
    \ENDFOR
    
    \FOR{each pixel \( (m, n) \) for \( c = 2, 3 \)}
        \STATE \( Y_{m,n,c} \gets \Delta^r X_{m,n,c} - \Delta^r X_{m,n,1} \)
    \ENDFOR
\ENDFOR
\RETURN \( Y \)
\end{algorithmic}
\end{algorithm}

\subsection{Bitmap Encoding}
\label{sec:bitmap}

The encodings given by the Hierarchical Projection Coding provide a new perspective of the original WSI images, where the range of the pixel values is largely shortened. 

Not every bit in the current byte stream carries meaningful data. For instance, leading zeros in certain encoding schemes may not represent valid information, with only the bits following the first '1' considered as containing useful data. We define these bits as effective bits, since only after encountering the first '1' do subsequent bits hold significance. For example, a sample sequence such as [1, -1, 2] would have most of its bit positions empty, with only the bits following the first '1' carrying relevant information.

The distribution of effective bits in the encodings produced by the Hierarchical Projection Coding algorithms exhibits two key characteristics: 
(a) higher-position bits contain less information than lower-position bits, and (b) the redundancy in low-position information increases.
Therefore, it is natural to reorganize the ordering of the delta-encoded information, such that bits from the same positions in each byte have higher predictability.
The high predictability enables us to transform the delta-encoded bit information to achieve a higher compression ratio, as shown in Fig.~\ref{fig:hp_encoding}. 

To gather the effective bits, we transpose the encodings produced by the Hierarchical Projection Coding into bitmaps, then group the bits from the same position across adjacent bytes and re-pack them into new byte representations. As shown in Fig.~\ref{fig:hp_encoding}, the proposed bitmap coding transforms the sample WSI matrix into a new format, revealing numerous repetitive patterns that can be collected using a dictionary-based compressor. In the next section, we will demonstrate that applying an additional dictionary collection step on top of this bitmap coding can further reduce the entropy of the sample WSI matrix.







\subsection{Dictionary Encoding}
As illustrated in Fig.~\ref{fig:hp_encoding}, the output of bitmap encoding contains clearly long patterns. Those patterns are raised by the grouping of effective bits and in-effective bits. Therefore, we incorporate a dictionary encoding method at the final stage of the compression process to achieve even lower entropy and a higher compression ratio. 
For the specific dictionary algorithm, we incorporate Lempel–Ziv–Welch (LZW) algorithm.
The LZW Algorithm is a dictionary-based lossless data compression method. Unlike Arithmetic Coding, for LZW compression, there is no need to know the probability distribution of each character in the text alphabet. This allows compression to be done while the message is being received. The main idea behind compression is to encode character strings that frequently appear in the text with their index in the dictionary. The first 256 words of the dictionary are assigned to extended ASCII table characters.

After dictionary coding, as shown in Fig.~\ref{fig:hp_encoding}, the entropy is further reduced from 2.94 to 2.54. This reduction is achieved because the dictionary encodes long patterns into shorter, repeated signals, transforming the matrix into a collection dominated by repetitive symbols. Additionally, by concentrating effective and ineffective bits separately, this final dictionary-based encoding step further compresses the information within the WSI image.

%
%
%







\section{Experiments}
In this section, we conduct extensive experiments on realistic WSI images to answer the following research questions:
\begin{itemize}
    \item \textbf{RQ1}: Can \ours achieve higher compression ratios than existing methods?
    \item \textbf{RQ2}: Can \ours achieve higher compression throughput compared to current methods?
    \item \textbf{RQ3}: What role does each technique in \ours play in improving compression performance?
\end{itemize}

\subsection{Experimental Setups}
\paragraph{Datasets} 

%
We selected five H\&E datasets and one IHC dataset as our benchmarks. The H\&E datasets include famous and well-known C16~\citep{DATASETc16}, C17~\citep{DATASETc17}, TCGA~\citep{DATASETtcga}, BNCB~\citep{xu2021predicting}, and DigestPath~\citep{da2022digestpath}. To further increase dataset diversity, we incorporated a recently emerging IHC~\cite{akbarnejad2023predicting} dataset named IHC4BC, alongside the more commonly used H\&E-stained datasets.

\paragraph{Baselines} Our selected baseline encompasses nearly all types of lossless compression algorithms, including image encoders, deep learning-based image encoders, entropy encoders, and dictionary-based encoders. Specifically, for image encoders, we selected PNG and TIFF; for deep learning-based image encoders, we chose ArIB-BPS. For entropy encoders, we included the well-known Huffman and arithmetic coding. As for dictionary-based encoders, which are of particular interest to us, we selected three options: GZip, Zstd-22(level 22), and LZMA.

\paragraph{Evaluation Metrics} For lossless compression, the compression ratio is the primary performance metric. Additionally, we also consider throughput. 
Lossless compression does not cause any information loss in the image. Therefore, unlike typical lossy methods, we do not measure PSNR and SSIM for the decoded image in our experiments.
However, we still use PSNR in Sec.~\ref{exp:emprical} to measure the similarity between bitmaps, aiming to demonstrate that the bitstream after bitmap coding has higher local similarity.

\subsection{Compression Performance on Gigapixel WSIs}
\label{sec:exp_comp_rate}
Tab.~\ref{tab:compression_results_datasets} presents the main compression results comparing \ours with the existing state-of-the-art compression methods.
We present a comparison of average compression ratios between six lossless baseline methods and our approach on three of the most
well-known datasets—C16, C17, and TCGA. As shown, \ours achieves a substantial improvement over all other lossless compression methods. Specifically, \ours improves compression rates by 100\% to 500\% compared to non-dictionary-based methods. For dictionary-based methods, our compression ratio surpasses Gzip by 100\%, and even outperforms the current best dictionary-based method, Zstd-22 (highest compression level of Zstd), by 70\% to 80\%. \ours’s compression improvements are consistent across the different datasets.

To further illustrate the characteristics of WSI data, we provide specific compression ratios for 10 sample WSI images, each approximately 2GB in size. The compression rates vary significantly across images, mainly due to two factors: (i) the substantial blank areas in WSI images, and (ii) variations in tissue density. For instance, the first sample in Fig.~\ref{fig:pipeline_overview} clearly contains large blank areas. Regardless of the image type, however, \ours consistently achieves exceptionally high compression ratios.

\begin{table*}[t]
\belowrulesep=0pt
\aboverulesep=0pt
\centering
\caption{Compression results for C16, C17, and TCGA datasets (without overall average column).}
\label{tab:compression_results_datasets_no_overall}
\resizebox{0.8\textwidth}{!}{
\begin{tabular}{l|*{10}{c}}
\toprule
\rule{0pt}{10pt}\multirow{3}{*}{Method} & \multicolumn{10}{|c}{\textbf{C16 Dataset}} \\
& \multicolumn{10}{|c}{Compression Ratios on Sample Images} \\
\cmidrule{2-11}
& Img1 & Img2 & Img3 & Img4 & Img5 & Img6 & Img7 & Img8 & Img9 & Img10 \\
\midrule
Huffman & $1.55$ & $3.93$ & $2.06$ & $4.51$ & $1.49$ & $1.90$ & $1.53$ & $1.53$ & $1.89$ & $1.61$ \\
PNG     & $2.97$ & $6.76$ & $5.12$ & $8.22$ & $2.82$ & $4.41$ & $2.73$ & $2.73$ & $4.06$ & $3.08$ \\
TIFF    & $2.38$ & $9.86$ & $6.15$ & $14.83$ & $2.22$ & $3.81$ & $2.38$ & $2.38$ & $4.11$ & $3.02$ \\
Gzip    & $2.40$ & $10.88$ & $6.56$ & $16.80$ & $2.23$ & $3.88$ & $2.40$ & $2.40$ & $4.58$ & $3.06$ \\
Lzma    & $3.44$ & $16.83$ & $9.25$ & $26.50$ & $3.14$ & $5.46$ & $3.37$ & $3.37$ & $6.58$ & $4.34$ \\
Zstd-22 & $3.12$ & $15.16$ & $8.54$ & $23.96$ & $2.82$ & $4.97$ & $3.62$ & $3.04$ & $11.89$ & $3.90$ \\
\ours   & $\mathbf{4.74}$ & $\mathbf{22.94}$ & $\mathbf{12.52}$ & $\mathbf{37.18}$ & $\mathbf{4.48}$ & $\mathbf{7.67}$ & $\mathbf{5.67}$ & $\mathbf{4.54}$ & $\mathbf{17.79}$ & $\mathbf{6.03}$ \\
\midrule
\rule{0pt}{10pt}\multirow{3}{*}{Method} & \multicolumn{10}{|c}{\textbf{C17 Dataset}} \\
& \multicolumn{10}{|c}{Compression Ratios on Sample Images} \\
\cmidrule{2-11}
& Img1 & Img2 & Img3 & Img4 & Img5 & Img6 & Img7 & Img8 & Img9 & Img10 \\
\midrule
Huffman & $3.55$ & $3.77$ & $5.09$ & $3.80$ & $5.91$ & $4.88$ & $4.30$ & $4.67$ & $4.38$ & $3.48$ \\
PNG     & $7.19$ & $8.34$ & $14.01$ & $8.48$ & $4.99$ & $13.39$ & $9.91$ & $12.59$ & $11.92$ & $4.59$ \\
TIFF    & $7.56$ & $8.23$ & $23.35$ & $9.04$ & $27.43$ & $18.89$ & $11.25$ & $17.47$ & $12.04$ & $6.55$ \\
Gzip    & $7.96$ & $8.75$ & $25.93$ & $9.17$ & $68.84$ & $19.31$ & $12.20$ & $17.68$ & $12.85$ & $7.47$ \\
Lzma    & $11.33$ & $12.21$ & $36.82$ & $12.99$ & $100.19$ & $26.92$ & $16.96$ & $25.25$ & $17.99$ & $10.53$ \\
Zstd-22 & $10.16$ & $10.89$ & $32.61$ & $11.61$ & $89.60$ & $23.96$ & $15.14$ & $22.48$ & $16.06$ & $9.41$ \\
\ours   & $\mathbf{15.82}$ & $\mathbf{16.56}$ & $\mathbf{50.68}$ & $\mathbf{17.84}$ & $\mathbf{136.15}$ & $\mathbf{37.31}$ & $\mathbf{23.51}$ & $\mathbf{35.56}$ & $\mathbf{24.05}$ & $\mathbf{14.91}$ \\
\midrule
\rule{0pt}{10pt}\multirow{3}{*}{Method} & \multicolumn{10}{|c}{\textbf{TCGA Dataset}} \\
& \multicolumn{10}{|c}{Compression Ratios on Sample Images} \\
\cmidrule{2-11}
& Img1 & Img2 & Img3 & Img4 & Img5 & Img6 & Img7 & Img8 & Img9 & Img10 \\
\midrule
Huffman & $2.11$ & $2.63$ & $2.01$ & $2.34$ & $1.39$ & $1.77$ & $1.73$ & $1.34$ & $1.50$ & $2.43$ \\
PNG     & $3.54$ & $4.41$ & $3.95$ & $2.95$ & $1.97$ & $1.95$ & $2.19$ & $2.11$ & $2.04$ & $4.81$ \\
TIFF    & $3.74$ & $4.93$ & $3.22$ & $5.03$ & $1.71$ & $2.17$ & $2.52$ & $1.69$ & $2.13$ & $6.44$ \\
Gzip    & $3.78$ & $5.01$ & $3.53$ & $5.11$ & $1.79$ & $2.52$ & $2.53$ & $1.69$ & $2.14$ & $6.56$ \\
Lzma    & $5.02$ & $6.66$ & $4.67$ & $6.80$ & $2.45$ & $3.31$ & $3.28$ & $2.28$ & $2.88$ & $9.18$ \\
Zstd-22 & $4.45$ & $6.00$ & $4.16$ & $6.06$ & $2.18$ & $2.97$ & $2.94$ & $2.04$ & $2.57$ & $7.96$ \\
\ours   & $\mathbf{6.26}$ & $\mathbf{8.02}$ & $\mathbf{6.02}$ & $\mathbf{8.35}$ & $\mathbf{3.69}$ & $\mathbf{4.59}$ & $\mathbf{4.30}$ & $\mathbf{3.17}$ & $\mathbf{3.85}$ & $\mathbf{11.16}$ \\
\bottomrule
\end{tabular}}
\end{table*}

\begin{table}
  \centering
  \caption{Comparison of our method with other traditional compressor on patches on various WSI datasets including C16, C17, TCGA, BNCB, and IHC4BC.}

  \label{tab:patch_comparison}
  \resizebox{0.45\textwidth}{!}{
  \begin{tabular}{c|ccccccc|}
    \hline
    \textbf{Method} & \textbf{C16} & \textbf{C17} & \textbf{BNCB} & \textbf{TCGA} & \textbf{IHC} &\\
    \hline 
    PNG & $1.64$ & $1.96$ & $1.46$ & $1.04$ & $2.75$ \\
    TIFF & $1.16$ & $1.15$ & $1.16$ & $1.15$ & $2.32$ \\\hdashline[1pt/1pt]
    Huffman & $1.08$ & $1.12$ & $1.12$ & $1.10$ & $1.45$ \\
    Arithmetic & $1.09$ & $1.12$ & $1.13$ & $1.13$ & $1.46$ \\\hdashline[1pt/1pt]
    Gzip & $1.15$ & $1.27$ & $1.15$ & $1.14$ & $2.00$ \\
    Zstd-22 & $1.24$ & $1.36$ & $1.17$ & $1.16$ & $2.48$ \\
    Lzma & $1.32$ & $1.46$ & $1.29$ & $1.32$ & $2.60$ \\
    \hline
    \textbf{WISE} & $\mathbf{2.82}$ & $\mathbf{3.20}$ & $\mathbf{1.63}$ & $\mathbf{2.50}$ & $\mathbf{3.85}$ \\
    \hline
  \end{tabular}}
\end{table}

We further investigate the compression performance on non-empty patches, as these regions pose the primary challenge in compressing WSIs. Here, we expand our dataset to include five source: C16 $(256\times256)$, C17 $(256\times256)$, TCGA $(256\times256)$, BNCB $(256\times256)$, and an IHC dataset $(1024\times1024)$. WISE consistently achieves significant compression improvements across different patch sizes, staining techniques, and datasets.

\subsection{Comparision with Neural-based Compressor}

We also compared our method with NN-based approaches and found that it consistently outperforms the state-of-the-art ArIB-BPS across all datasets, with improvements of 7.1\%, 17.6\%, and 5\% on C16, C17, and TCGA, respectively. Additionally, we compared the efficiency of our method with NN-based compressors. This experiment was conducted on a CPU to ensure a fair comparison. As shown, the NN-based method is extremely slow, with a throughput of only 18.73 KB/s, whereas our method achieves a throughput of 15.1 MB/s, suitable for practical use. Furthermore, the NN-based method requires 3.81 GB of memory even for compressing a $256\times256$ patch, while WISE only needs 0.1 GB.

\begin{table}
  \centering
  \caption{Comparison of WISE with Neural-Network compressor on patches from various WSI datasets including C16, C17, TCGA, BNCB, and IHC4BC. Throu. stands for throughput and Mem. stands for memory usage. }
  \label{tab:patch_comparison}
  \resizebox{0.45\textwidth}{!}{
  \begin{tabular}{c|cccc|cc}
    \hline
    \textbf{Method} & \textbf{C16} & \textbf{C17}  & \textbf{TCGA} & \textbf{IHC} & \textbf{Throu.} & \textbf{Mem.} \\
    \hline
    ArIB-BPS & $2.63$ & $2.72$ & $2.38$ & $3.81$ & 18.73KB/s & 3.81GB \\
    \textbf{WISE} & $\mathbf{2.82}$ & $\mathbf{3.20}$ & $\mathbf{2.50}$ & $\mathbf{3.85}$ & 15.1MB/s & 0.1GB \\
    \hline
  \end{tabular}}
\end{table}



\subsection{Framework Effectiveness}

\begin{figure}[t]
  \centering
\includegraphics[width=0.48\textwidth,angle=0]{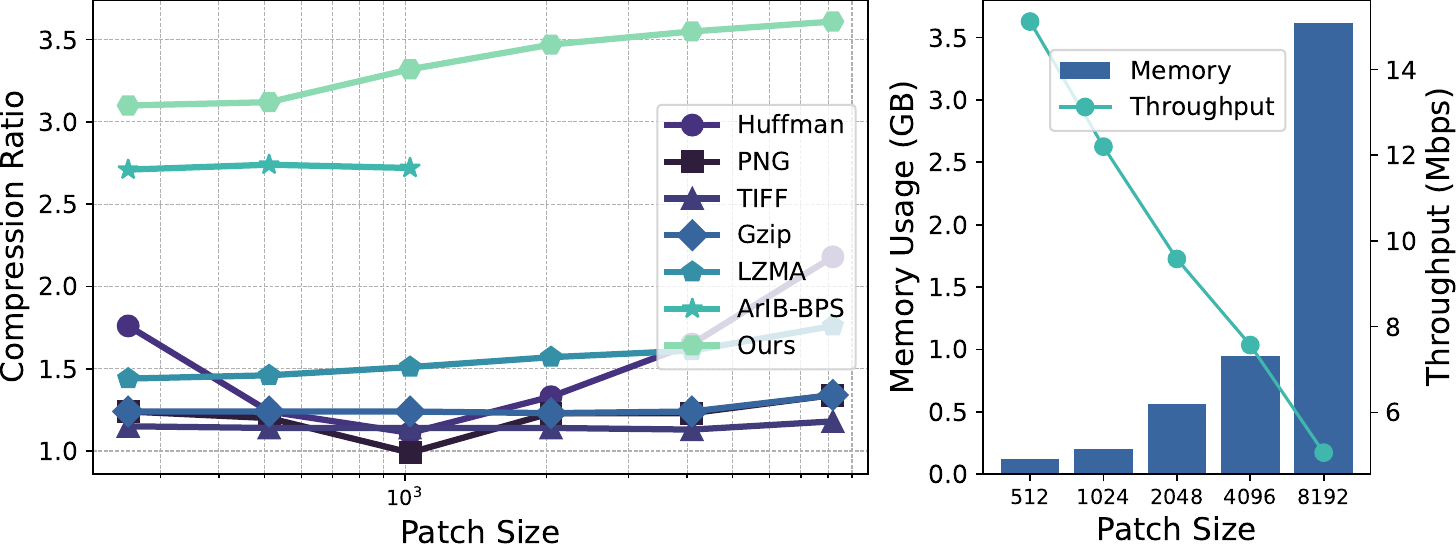}
  \caption{Left: Compression ratio variations on different image sizes. Right: \ours's memory usage and throughput on different image sizes.}
  \label{fig:memory}
\end{figure}

To further evaluate the practical applicability of our method, we explore its throughput and memory usage across different input image sizes in this section. All experiments were conducted using single-threaded Python code. We focus primarily on the compression efficiency of the informative areas, as the compression of empty areas has minimal impact. While the search algorithm for empty areas is important, it falls outside the scope of this paper, which is focused on compressing the informative regions.
As observed, both throughput decreases and memory usage increases with the input size. This is primarily due to the LZW algorithm's dictionary collection process, where memory usage grows as the input size increases, driven by the increased complexity of the search. However, this is accompanied by an improvement in compression ratio.
It is worth noting that when the input size reaches $8192\times8192\times3$, our method achieves a memory usage comparable to ArIB-BPS, but with a significantly lower memory footprint of only 3 GB.


\subsection{Ablation Study}

\paragraph{Effectiveness of Proposed Techniques} We conducted ablation experiments on C16, C17, and TCGA to demonstrate the effectiveness of each proposed technique.
%
The baseline shown in the table is a basic dictionary-based algorithm, LZMA. Results indicate that applying LZMA alone is not highly effective, even though, as discussed in the previous section, LZMA provides a certain level of compression compared to other lossless algorithms. Medcomp improves the compression ratio of the dictionary-based method to 113.5\%$\sim$130.2\%. Specifically, the proposed Hierarchical encoding achieves relative improvements over the baseline by 69.6\%, 76.9\%, and 73.5\% on C16, C17, and TCGA, respectively. Furthermore, the proposed bitmap encoding can further increase the compression ratio to 125.6\%, 130.2\%, and 113.5\%.

\begin{table}
  \centering
  \caption{Ablation Study on various WSI datasets including C16, C17, TCGA.}
  \small
  \label{tab:comparison}
  \begin{tabular}{c|ccc}
    \hline
    \textbf{Method} & \textbf{C16} & \textbf{C17} & \textbf{TCGA}\\
    \hline
    Baseline & 1.25 & 1.39 & 1.17 \\
    +Hierarchical Linear Projection & $2.12$ & $2.46$ & $2.03$ \\
    +Bitmap Encoding & $2.82$ & $3.20$ & $2.50$ \\
    \hline
  \end{tabular}
\end{table}

\paragraph{Compression ratio under different input image sizes.} In Fig.~\ref{fig:memory}, we show the variation in compression ratios for six baselines and our method across different input image sizes. The ArIB-BPS, a neural network-based approach, was tested on a single 24GB Nvidia-4090 GPU. Due to memory constraints, it was limited to processing images of size $1024\times1024$ and smaller. As observed, the compression ratio for images smaller than $1024\times1024$ remains largely unchanged.
Another key observation is that all dictionary-based compression methods show improved compression ratios as the input size increases. This is intuitive, as dictionary-based methods can better exploit previously collected patterns when more input data is available. In contrast, the compression ratios for entropy-based methods are less consistent and do not show the same trend.


\subsection{Empirical Study}
\label{exp:emprical}

\begin{figure}[t]
  \centering
\includegraphics[width=0.45\textwidth,angle=0]{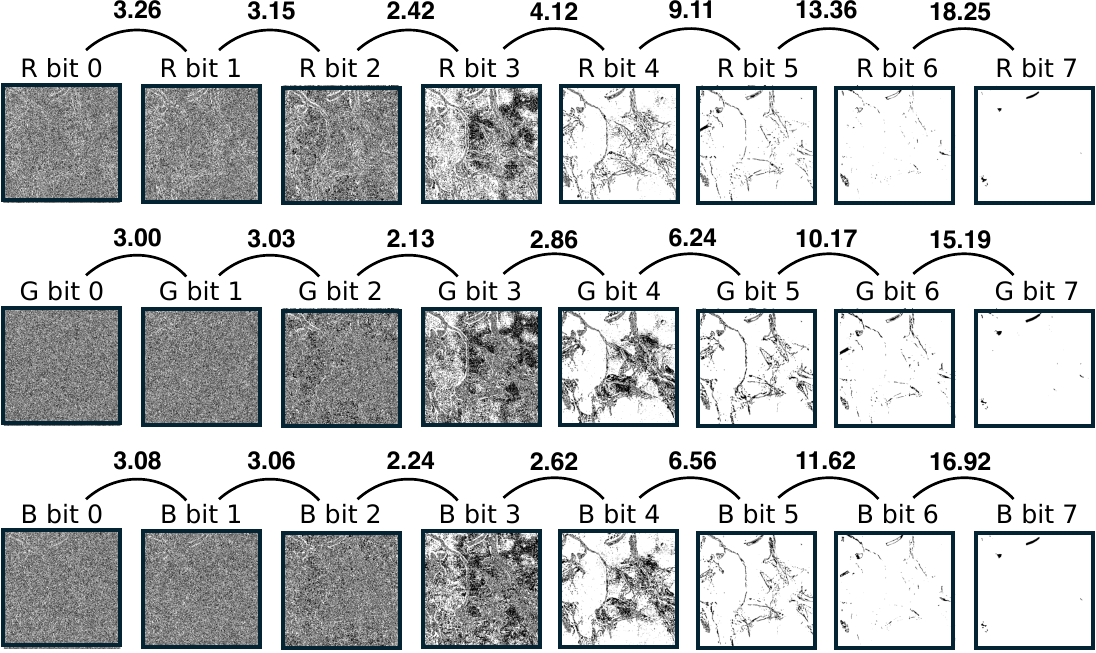}
  \caption{PSNR of bitmaps before Bitmap Coding. As illustrated, bitmaps in higher bit positions have larger empty areas and higher PSNR values, indicating they contain less information..}
  \label{fig:psnr1}
\end{figure}

\begin{figure}[t]
  \centering
\includegraphics[width=0.45\textwidth,angle=0]{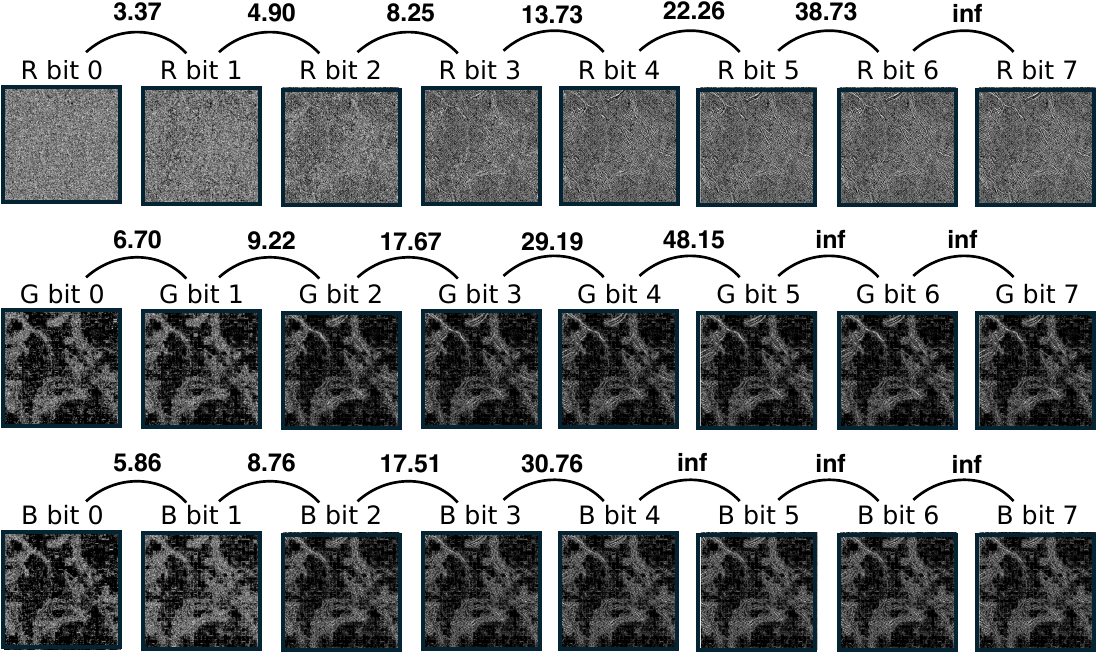}
  \caption{PSNR of bitmaps after Bitmap Coding. As illustrated, the PSNR between bitmaps increases significantly, indicating more repetitive patterns among bitmaps, which enhances the efficiency of dictionary coding.}
  \label{fig:psnr2}
\end{figure}

We present an empirical study on a specific WSI patch to validate several conclusions discussed earlier from a different perspective.
Figs.~\ref{fig:psnr1} and Fig~\ref{fig:psnr2} illustrate the similarity between the original WSI patch and its bitmap representation after delta coding, with similarity measured using PSNR. Bitmap transposition improves the efficiency of dictionary collection.
From Fig.~\ref{fig:psnr1}, it can be observed that after performing a bitmap split on the original image, the bitmap similarity for each channel is around 3 to approximately 18. Moreover, the similarity between lower-bit bitmaps is lower, while the similarity between higher-bit bitmaps is higher. 
%
Next, we examine Fig~\ref{fig:psnr2}, which shows the similarity between the bitmaps of the sample patch after applying the proposed hierarchical linear coding.
The first conclusion from this figure is that the similarity between the bitmaps after hierarchical linear coding significantly increases compared to the original patch, rising from 3 to 48. In fact, some higher-bit bitmaps reach an infinite PSNR value, as they have no information variance and are identical, confirming that "the range of pixel values is largely shortened by hierarchical linear coding" as stated in Sec.~\ref{sec:bitmap}. Hierarchical linear coding further concentrates information in the lower bit positions.
The second conclusion is that, due to the high similarity between the bitmaps after hierarchical linear coding, they are more easily grouped into similar patterns, making them more compressible. Indeed, during the experimental process, we observed a similar phenomenon in the dictionary coding stage. The collected patterns across different channels and bitmaps showed high similarity.


\section{Conclusions}
In this work, we explored applying compression techniques to WSI images. Through the study of existing compression techniques on WSI images, we identified several unique challenges in compressing WSI images, including how to deal with information sparsity and information irregularity. To resolve those challenges, we developed a simple yet effective WSI specialized compressor called \ours. \ours is built upon a hierarchical encoding strategy and a dictionary-based compression method, which can be further combined with various entropy-based compression methods to further improve the compression ratios. Extensive experiments on realistic WSI benchmarks show that \ours achieves up to $136\times$ and on average $36\times$ compression over prior methods, while retaining low decoding complexity and preserving critical diagnostic content, making it highly practical for deployment in real-world digital pathology workflows across diverse tissue types and scanning conditions.
{
    \clearpage
    \small
    \bibliographystyle{ieeenat_fullname}
    \bibliography{main}
}


\end{document}